\documentclass[aps,pre,amsmath,amssymb,preprint,superscriptaddress]{revtex4-2}
\usepackage{float}
\usepackage{graphicx}
\usepackage{xcolor}
\usepackage[export]{adjustbox}
\usepackage[utf8]{inputenc}
\usepackage{bm}
\usepackage{hyperref}
\usepackage{indentfirst}
\usepackage{natbib} 

\begin{document}
\title{Separation of interacting active particles in an asymmetric channel}

\author{Ankit Gupta} 
\email{ankitgupta@kgpian.iitkgp.ac.in}
\affiliation{Department of Physics, Indian Institute of Technology Kharagpur, Kharagpur 721302, India}

\author{P. S. Burada }
\email{Corresponding author: psburada@phy.iitkgp.ac.in}
\affiliation{Department of Physics, Indian Institute of Technology Kharagpur, Kharagpur 721302, India}

\date{\today}

\begin{abstract}

We study the diffusive behaviour of interacting active particles (self-propelled) with mass $m$ in an asymmetric channel. The particles are subjected to an external oscillatory force along the length of the channel. In this setup, particles may exhibit rectification. In the absence of interaction, the mean velocity $\langle v \rangle$ of the particles shows a maximum at moderate $m$ values. It means that particles of moderate $m$ have higher velocities than the others. However, by incorporating short-range interaction between the particles, $\langle v \rangle$ exhibits an additional peak at lower $m$ values, indicating that particles of lower and moderate $m$ can be separated simultaneously from the rest.
Furthermore, by tuning the interaction strength, the self-propelled velocity, and the parameters of the oscillatory force, one can selectively separate the particles of lower $m$, moderate $m$, or both. Empirical relations for estimating the optimal mass as a function of these parameters are discussed. These findings are beneficial for separating the particles of selective $m$ from the rest of the particles. 
\end{abstract}

\maketitle

\section{Introduction}
\label{sec:intro}

Microorganisms self-propel in a fluid environment, and synthetic active particles use their environment's free energy to convert it into sustained motion. 
The processes behind such an active motion have been widely investigated \cite{Ramaswamy, Lauga_2009, Zottl_2016, Marchetti2013, Caprini2021, Elgeti2015, Khatri2022, Khatri2023}. A new field known as the active matter has been emerging, focusing on the physical features of propulsion, processes, and motility-induced collective behaviour of more identical entities. Recently, more attention has been directed toward considering inertia's influence on active particle movement. 
Active particles differ from passive particles in their degree of freedom. Passive particles move randomly due to thermal fluctuations, whereas active particles move persistently and autonomously due to internal driving forces. 
Consequently, active particles collectively exhibit peculiar behaviours different from passive particles. Biological and synthetic active agents often function in thick fluids where inertia usually isn't significant. However, there are cases where inertia can't be ignored. Lately, researchers have been focusing more on understanding how inertia influences the movement of active particles \cite{Scholz2018,Sprenger2022}. This interplay between active forces and particle inertia can lead to intriguing behaviours and collective effects in certain systems. \cite{Arkar2021,Bechinger2016,Kolmakov2021} 
Furthermore, the complex interactions among the active particles can significantly influence the behaviour of active matter systems, leading to various patterns and structures \cite{Ramaswamy, Lauga_2009, Marchetti2013, Elgeti2015}.  
The interactions between the active particles are complex and arise due to various processes, 
e.g., mechanical collisions, hydrodynamic interactions, and lubrication forces. 

When active or passive particles move in confined structures such as porous media \cite{Wang}, microfluidic channels\cite{Matthias2003}, and living tissues \cite{Bressloff}, the geometry of the channel controls their diffusive behaviour. 
These confined geometries create an entropic barrier that can affect particle diffusion. This unique property makes them useful for various applications such as chemical processing, separation techniques, and catalysis. This has led to the development of efficient filtering materials, targeted drug delivery systems, and lab-on-a-chip devices for chemical analysis and medical diagnostics. Overall, the potential applications of confined geometries are vast and continue to be explored as research in this field progresses \cite{D_Reguera, Burada_2009, G_Schmid}.  
For instance,  industrial, biomedical, and clinical applications rely on separating and sorting small particles, e.g., wastewater purification, blood sample preparation, and disease diagnosis \cite{jin}. 

Separation techniques in research often hinge on the response of particles to external stimuli such as gradients or fields \cite{MacDonald2003}. The behaviour of these particles, whether they drift or diffuse, is influenced by various factors, including mass, size, shape, and charge. One can effectively separate and study individual particles by understanding and manipulating these properties. There is an overwhelming need to separate mesoscopic particles from mixtures in laboratory research and industrial applications according to their physical characteristics \cite{D_Reguera, Haenggi2009, Volkmuth1992, Bogunovic2012}. Separating mesoscopic constituents, such as malignant tumor cells or nanoparticles, can be challenging due to their small size and similarity to other particles. Consequently, innovative methods have emerged in the biomedical and technological domains to tackle these obstacles head-on, including developing micro-fabricated sieves or membranes, centrifugation, filtration, and microfluidics. However, these methods can be limited by factors such as the complexity of the sample, the need for high throughput, and the sensitivity of the particles to the separation conditions. 
Research in this area continues to be an active area of study to develop more efficient and effective methods for separating mesoscopic particles.
In mesoscopic particle separation, the emphasis on mass-based separation takes precedence over size-based methods due to its ability to discern specific particles according to their mass. This approach yields crucial insights into the potential involvement of these particles in disease initiation and propagation, as supported by notable studies \cite{Mukhopadhyay2018, Slapik2019, Lindner1999, Borromeo2007}. 
For example, in cancer research, the mass of cancer cells can be different from healthy cells, and this difference in mass can be used to identify and isolate cancer cells \cite{Suresh} for further study. Mass-based separation methods can detect and isolate specific proteins or pathogens in a sample associated with a specific disease. 

This work aims to study a mechanism for separating interacting active particles diffusing in an asymmetric channel in the presence of an oscillatory force \cite{Walther2013,Koleoso2020,Kalinay2014}. Separating particles based on their response to external force is a complex process involving multiple factors. 
We are interested in understanding how the reflection boundary conditions at the channel walls, particle interaction, and channel aspect ratio influence the separation of active Brownian particles in this constrained environment. With this study, we can gain a deeper understanding of the separation process of active particles and develop new effective methods for particle separation.
This article is organized as follows. 
In Sec.~\ref{sec:model}, we introduce the Langevin model to describe the dynamics of the active particles in a two-dimensional asymmetric channel. 
The transport characteristics of active parameters as a function of the interaction strength, the channel parameters, and the external oscillatory force are discussed in Sec.~\ref{sec:transport}. 
Sec~\ref{sec:discussion} and ~\ref{sec:conclusions} are devoted to the discussion and the main conclusions, respectively.

\section{MODEL}
\label{sec:model}

\begin{figure}[ht]
\centering
\includegraphics{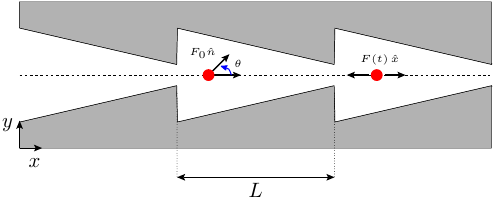}
\caption{Schematic illustration of a 2D asymmetric channel, described by Eq.~\ref{eq:channel}, with the periodicity $L$ confining the motion of an active Brownian particle of mass $M$. 
The particle propels with a self-propelled velocity $v_0 = F_0/\eta$ and is subjected to oscillatory force $F(t)$. 
Eq.~\ref{eq:interaction} describes the interaction between the particles.}
\label{fig:channel}
\end{figure}

Consider the dynamics of an active Brownian particle of mass $M$ suspended in a thermal bath and constrained by a two-dimensional (2D) asymmetric triangular channel, as shown in figure \ref{fig:channel}. 
An oscillatory force is applied on the particle along the length of the channel, i.e., $x$ direction.
In the presence of interaction between the particles, the dynamics of the particle is described by 
the Langevin equations as
\begin{align} 
M\frac{d^2\vec{r}}{dt^2} &= - \eta\frac{d\vec{r}}{d t} + F_0 \hat{n} + \vec{F}_{\text{int}} + F(t) \hat{x} + \sqrt{\eta k_B T} \, \vec{\xi}(t) \\
\frac{d\vec{\theta}}{d t} & = \sqrt{\eta k_B T} \, \vec{\chi}(t)\,,
\end{align}
where $\vec{r} $ is the position vector of the particle in 2D, 
orientation of the particle $\hat{n} = (\cos \theta , \sin \theta)$, 
$\theta$ is the angle relative to the $x$ axis,  
$v_0$ is the self-propelled velocity of the particle with the corresponding active force $F_0 = \eta \,v_0$, $\eta$ is the friction coefficient, 
$k_B$ is the Boltzmann constant, and $T$ is the temperature of the surrounding medium. 
The oscillatory force is described by $\vec{F}(t) = A \sin(\Omega t)\,\hat{x}$, 
where $A$ is amplitude and $\Omega$ is driving frequency. 
The thermal fluctuations due to the coupling of the particle with the surrounding medium are modelled by a zero-mean Gaussian white noise $\vec{\xi(t)}$ and $\vec{\chi}(t)$, obeying the fluctuation-dissipation relation 
$\langle \xi_i(t)\xi_j(t') \rangle = 2\delta_{ij} \, \delta(t-t')$ for $i,j = x,y$ (similarly for $\chi(t)$).
Note that in this study, we have considered the strength of noise functions to be the same for the translational and rotational thermal fluctuations for simplicity. However, they may generally be different depending on the system under consideration.

A two-dimensional asymmetric and spatially periodic channel is described by its half-width as
\begin{equation}
\label{eq:channel}
w_u(x) =
\begin{cases}
w_\mathrm{min}, & {x = 0} \\
w_\mathrm{max} - (w_\mathrm{max} - w_\mathrm{min}) \frac{x}{L}, & {0 < x \le L}\,,
\end{cases}
\end{equation}
where $w_\mathrm{max}$ and $w_\mathrm{min}$ are the maximum and minimum half-widths of the channel, respectively, and $L$ represents the periodicity of the channel. The ratio of these two widths defines the dimensionless aspect ratio given by
\begin{equation}
\label{eq:aspect}
\epsilon = \frac{w_\mathrm{min}}{w_\mathrm{max}} \,,  \ \ \ \ {0 < \epsilon \le 1}.
\end{equation}
The upper wall of the channel is defined as $ w_u(x)$ = - $w_l(x)$ due to the symmetry about the channel direction ($x$ axis). As a result,  $2w(x)$ = $w_u(x)$ - $w_l(x)$  gives the local width of the channel.
The dynamics of the self-propelled particle at the channel walls are modelled as follows. 
The particles cannot pass through the rigid walls of the channel. However, they are free to rotate and slide within the channel. The translational velocity $\dot{\vec{r}}$ is elastically reflected, and the rotational angle $\theta$ is unchanged during the collision (sliding reflecting boundary condition) \cite{Khatri2019, Ghosh2013, Reichhardt2017}.

The short-range interaction force on a particle $i$ due to its neighbours can be calculated using the lubrication theory \cite{OtextquotesingleNeill1967} as $\vec{F}_\mathrm{int} =  \sum_{j \,(i\neq j} \sigma_{ij} (\cos \alpha_{ij} \,\hat{x} + \sin \alpha_{ij} \,\hat{y})/d_{ij}\,$,
where the sum is taken over all its nearest neighbours, $\sigma_{ij}$ is the interaction strength between the particles $i$ and $j$, $d_{ij}$ is the corresponding distance between the particles, $\alpha_{ij}$ is the angle that $d_{ij}$ makes with the channel axis ($x$-axis), and $\hat{x}$ and $\hat{y}$ are the unit vectors along the $x$ and $y$ directions, respectively. 
Note that $\sigma_{ij} > 0$ (or $\sigma_{ij} < 0$) means the particles are repelling (or attracting) each other. However, in the limit of the low density of the particles, this interaction force can be approximated as (see Refs.~\cite{Khatri2019})
\begin{equation}
\label{eq:interaction}
\vec{F}_\mathrm{int} =  k \,(\cos \alpha \,\hat{x} + \sin \alpha \,\hat{y})\,,
\end{equation}
where $k$ denotes the total strength of the interaction force on a particle due to its neighbours 
and $\alpha$ is a random variable that can have values between $0$ to $2\pi$.
Eq.~\ref{eq:interaction} is like the mean force acting on a particle due to its neighbours. 
In the following,  Eq.~\ref{eq:interaction} is considered for the particle interaction. 

To have a dimensionless description, we scale all lengths by the periodicity of the channel $L$ and time by $\tau  = \eta L^2/(k_B T)$, which represents the typical diffusion time \cite{Khatri2021}.
The dimensionless Langevin equation reads
\begin{align} 
\label{eq:lan-dl}
m\frac{d^2\vec{r}}{dt^2} & = - \frac{d\vec{r}}{d t} + \ f_0 \, \hat{n}  + \ \vec{f}_\mathrm{int} + f(t) \,\hat{x} + \vec{\xi}(t) \\
\label{eq:lan-dlth}
\frac{d\vec{\theta}}{d t} & = \vec{\chi}(t) \,,
\end{align}
where $m = \tau_0/\tau = M k_B T/(\eta^2 L^2)$, $\tau_0 = M/\eta$ is the typical time of velocity relaxation for the active Brownian particle. As a result, the dimensionless mass $m$ is impacted by the particle's physical mass $M$ and the friction coefficient $\eta$, thermal energy $k_B T$, and channel periodicity $L$. $f = F_0 L/(k_B T)$ is the dimensionless active force. 
$a = A L/(k_B T)$ and $\omega = \Omega \tau$ are the dimensionless amplitude and frequency of the oscillatory force, respectively. 
$\vec{f}_\mathrm{int} = \vec{F}_\mathrm{int} L/(k_B T) = k L/(k_B T)$ is the dimensionless interaction force. 
In the rest of the paper, we use dimensionless variables.

\section{Transport characteristics}
\label{sec:transport}

As the particles are confined in the transversal direction of the channel, 
we can compute the mean velocity $\langle v \rangle$ and the effective diffusion coefficient $D_{eff}$ along the length of the channel ($x$ direction). Unfortunately, we cannot calculate them analytically by solving the corresponding Fokker-Planck equation with the reflecting boundary conditions at the channel walls.  
For this reason, we rely on numerical simulations. 
Thus, $\langle v \rangle$ and $D_{eff}$ can be calculated using Brownian dynamics simulations performed by solving the Langevin equation Eqs.~\ref{eq:lan-dl} \,\&~\ref{eq:lan-dlth} using the standard stochastic Euler algorithm over $10^4$ trajectories with reflection boundary conditions at the channel walls (see Refs. \cite{Khatri2019, Khatri2022}). 
In the long-time limit, $\langle v \rangle$ and $D_{eff}$ of the ensemble of the particles along the $x$ direction can be calculated as 
\begin{align}
\label{eq:meanvel}
\langle v \rangle & = \lim_{t\to\infty} \frac{\langle x(t) \rangle}{t} \,,\\
D_{eff} & = \lim_{t\to\infty} \frac{\langle x^2(t) \rangle - \langle x(t) \rangle^2}{2 \, t}\,.
\end{align}

\subsection{Effect of particle interaction}

\label{sec:interaction}

\begin{figure}[ht]
\centering
\includegraphics{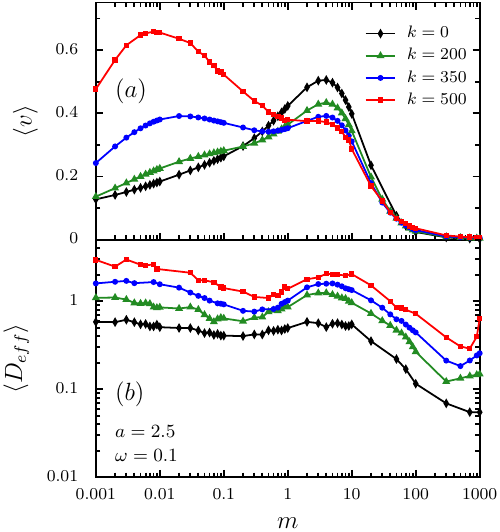}
\caption{ 
(a) The mean velocity $\langle v \rangle$ and (b) the effective diffusion coefficient $D_{eff}$ as a function of the mass $m$ of the active particles for different strengths of the interaction $k$ between the particles. 
The amplitude $a$ and the frequency $\omega$ of the oscillatory force are set to 
$\omega = 0.1$ and $a = 2.5$. The geometry of the channel is described by Eq.~\ref{eq:channel} with the channel aspect ratio $\epsilon = 0.1$.}
\label{fig:vel-deff}
\end{figure}

Fig.~\ref{fig:vel-deff} depicts the behaviour of $\langle v \rangle$ and $D_{eff}$ of active particles concerning their mass $m$ for various interaction strengths $k$. $\langle v \rangle$ exhibits  rectification in the channel direction. In the present case, $\langle v \rangle$ is positive because of the chosen channel geometry.  
The oscillatory force breaks the thermodynamic equilibrium and induces directed transport due to the asymmetry of the channel. 
In the absence of particle interaction, $\langle v \rangle$ shows a maximum at moderate $m$ values.  
It means that particles of moderate $m \,(\sim 10)$ have higher velocities than the others. 
A similar behaviour was reported in the case of passive particles \cite{Khatri2021,Ai2009}. 
The corresponding $D_{eff}$ also shows a similar trend. 
For particles of higher $m$, the inertia of the particles starts to dominate over the strength of the oscillatory force. As a result, both $\langle v \rangle$ and $D_{eff}$ decay quickly. 
In the other limit, i.e., $m \to 0$, viscous forces due to the surrounding medium dominate, which leads to a decrease in $\langle v \rangle$ with decreasing $m$. However, $D_{eff}$ of the active particles can be finite.

In the presence of interaction, $\langle v \rangle$ exhibits an additional peak at lower $m$ values. Indicting that lower and moderate $m$ particles can move faster than the rest. This behaviour can be effectively controlled by modifying the strength of the particle interaction $k$. $D_{eff}$ follows a similar behaviour.  
This bimodal pattern in $\langle v \rangle$ with respect to $m$ is solely due to the interaction between the particles. 
Note that increasing $k$ implies an increase in the interaction radius among the particles. Particles of lighter mass diffuse faster than the heavier ones (see Fig.~\ref{fig:vel-deff}(b)). 
It leads to higher rectification, 
as shown in Fig.~\ref{fig:vel-deff}(a). 
The system is nearly overdamped for particles of lighter mass $m$, and the effect of inertia is weak. 
As a result, $\langle v \rangle$ increases with $k$ in this limit. 
However, the peak height of $\langle v \rangle$ at moderate $m$ values decreases as $k$ increases. 
In this case, although the diffusivity of the particles increases with $k$, due to the inertia of the particles, the net displacement decreases. 

\begin{figure}[h!]
\centering
\includegraphics{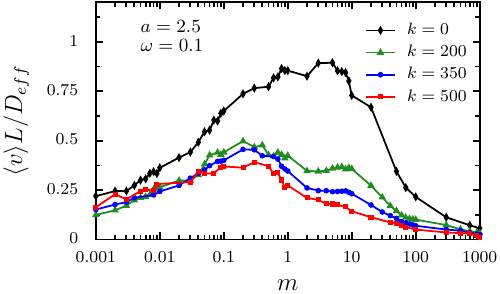}
\caption{$\langle v \rangle$L/$D_{eff}$ as a function of the mass $m$ of the active particles for different strengths of the interaction $k$ between the particles. 
The amplitude $a$ and the frequency $\omega$ of the oscillatory force are set to 
$\omega = 0.1$ and $a = 2.5$, with the channel aspect ratio $\epsilon$ = 0.1 and $L = 1$.}
\label{fig:vel-deff2}
\end{figure}

To ensure successful particle separation, attaining a peak in the mean velocity within the target mass and concurrently minimizing diffusivity lead to the formation of a precisely focused collimated beam of particles. In this context, investigating the efficiency parameter  $\frac{\langle v \rangle L}{D_{eff}}$ concerning mass $m$ is vital. 
Fig.~\ref{fig:vel-deff2} depicts the behaviour of $\langle v \rangle L/D_{eff}$ as a function of particle mass $m$ for various strengths of particle interaction $k$. 
Remarkably, the plot exhibits a bimodal structure (similar to Fig.~\ref{fig:vel-deff}(a)) for optimal values of $k$. This bimodal structure can signify an optimal regime where both high mean velocity and low diffusivity align, resulting in enhanced particle separation and a more focused particle beam.

\subsection{Variation of frequency and amplitude of the oscillatory force}
\label{sec:frequency}
\begin{figure}[ht]
\centering
\includegraphics{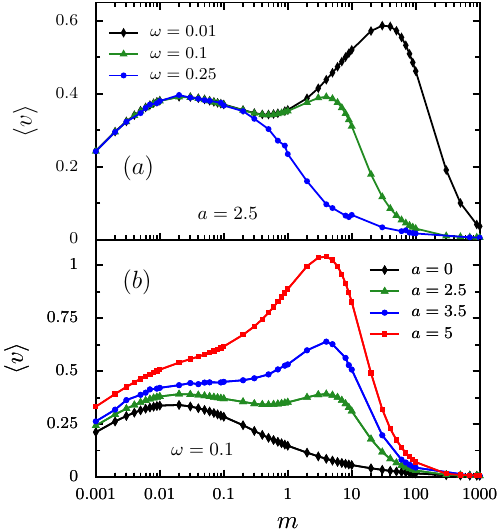} 
\caption{The mean velocity $\langle v \rangle$ as a function of the mass $m$ of the interacting particles for different strengths of frequency $\omega$ (a) and amplitude $a$ (b) of the oscillatory force. 
The geometry of the channel is described by Eq.~\ref{eq:channel} 
with the channel aspect ratio $\epsilon = 0.1$. 
Here, the particle interaction strength $k = 350$.}
\label{fig:a-omega}
\end{figure}

Fig.~\ref{fig:a-omega} shows the behaviour of $\langle v \rangle$ as a function of $m$ of the interacting particles for different frequency strengths $\omega$ and amplitude $a$ of the oscillatory force. 
The bimodal pattern collapses into a single peak as $\omega$ increases 
(see Fig.~\ref{fig:a-omega}(a)). 
Also, as $\omega$ increases, the peak shifts to the lower $m$ region. 
Furthermore, the peak value of $\langle v \rangle$ at moderate $m$ decreased with increasing $\omega$. 
At low frequencies, the particles have enough time to respond to the driving force and be able to move along the oscillatory force direction  and maintain relatively stable trajectories, resulting in two distinct peaks in $\langle v \rangle$. At higher frequencies, particles of moderate $m$ experience more rapid and irregular movements due to inertia. 
As a result, the peak at the moderate $m$ values decays quickly with increasing $\omega$ (see Fig.~\ref{fig:a-omega}(a)). 
However, the lighter particles respond to the oscillatory force and exhibit a peak in the small $m$ regime. 
As mentioned above, in the limit $m \to 0$, $\langle v \rangle$ is low. 

In the absence of the oscillatory force, i.e., $a = 0$, particles of lighter mass show rectification along the channel direction due to the chosen asymmetric channel shape (see Fig.~\ref{fig:a-omega}(b)). 
Here, the peak at the lower $m$ values is due to the active nature of the particles and the interaction between them. 
Since active particles can self-propel and have rotational degrees of freedom, they can diffuse more than passive particles. Due to this reason, active particles exhibit a more pronounced peak in $\langle v \rangle$. 
However, as $a$ increases, i.e., in the presence of oscillatory force, $\langle v \rangle$ shows a bimodal behaviour. 
For higher values of $a$, particles of moderate $m$ move faster than those of lighter $m$. 
As a result, $\langle v \rangle$ exhibits a single peak. 
From these observations, it is evident that the behaviour of $\langle v \rangle$ can be effectively tuned by modifying the parameters of the oscillatory force. As reported for the case of passive particles, the optimal values of $m$, i.e., the values of moderate $m$ correspond to the peak value of $\langle v \rangle$, can be estimated as $m_{op} \sim (a\,\omega^2)^{-0.4}$ \cite{Khatri2021}.
However, in the presence of the interaction, the optimal values of $m$ vary as $k^2$ for moderate $m$ values at fixed $a$ and $\omega$. Thus, in terms of $a$, $\omega$, and $k$, $m_{op} \sim (a\,\omega^2\, k^2)^{-0.4}$. 

\subsection{Variation of channel aspect ratio}
\label{sec:aspect-ratio}

\begin{figure}[ht]
\centering
\includegraphics{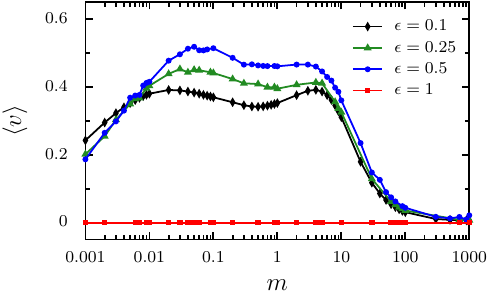}
\caption{The average velocity $\langle v \rangle$ as a function of the mass $m$ of active particles for different values of channel aspect ratio $\epsilon$. The geometry of the channel is described by Eq.~\ref{eq:channel}.
The amplitude $a$ and the frequency $\omega$ of the oscillatory force are set to 
$\omega = 0.1$ and $a = 2.5$. Here, the particle interaction strength $k = 350$.}
\label{fig:aspect}
\end{figure}

Figure ~\ref{fig:aspect} illustrates the behaviour of $\langle v \rangle$ as a function of $m$ for 
various channel aspect ratios (Eq.~\ref{eq:aspect}).  
For $\epsilon < 1$, geometric effects significantly control the transport characteristics of the particles due to modulation in the channel's shape. As mentioned, these geometric effects lead to entropic barriers, which dictate particle motion. Note that smaller $\epsilon$ corresponds to a highly confined channel, and higher $\epsilon (\sim 1)$ corresponds to a less confined channel \cite{Burada_2009}. In the presence of interaction between the particles, for $\epsilon < 1$, $\langle v \rangle$ shows a bimodal behaviour. However, as $\epsilon$ increases, the peak values at lower and moderate values of $m$ are more pronounced. It is because, as $\epsilon$ increases, the bottle-neck width increases, which leads to an increase in $\langle v \rangle$. 
For $\epsilon = 1$, which corresponds to a flat channel, as expected, particles do not show any rectification leading to $\langle v \rangle = 0$.

\subsection{Effect of self-propelled velocity}

\begin{figure}[ht]
\centering
\includegraphics{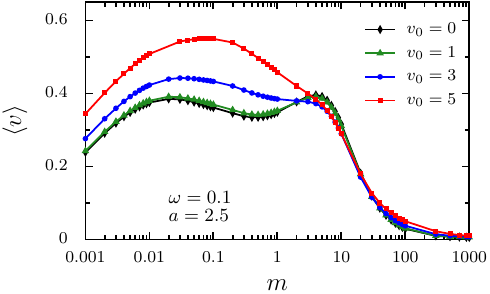}
\caption{
The average velocity as a function of the mass of active particles for different strengths of 
the self-propelled velocity $v_0$.
The particle interaction strength $k = 350$. 
The geometry of the channel is described by Eq.~\ref{eq:channel} with an aspect ratio $\epsilon = 0.1$.}
\label{fig:v0}
\end{figure}

Figure ~\ref{fig:v0} depicts the behaviour of $\langle v \rangle$ as a function of $m$ for various strengths of the self-propelled velocity $v_0$.
For $v_0 = 0$, i.e., for passive particles, $\langle v \rangle$ shows bimodal behaviour in the presence of particle interaction. 
However, as $v_0$ increases, the peak height of $\langle v \rangle$ at lower $m$ increases. 
It means particles of lower $m$ show higher rectification than the rest. 
Note that, in general, particles exhibit rectification while diffusing in asymmetric channels, even in the absence of oscillatory force. It is because particles shuffling between the cells of the asymmetric channel occurs due to the random force arising from thermal fluctuations resulting from the coupling of the particle with the surrounding medium. Additionally, for active particles, the random force can also be attributed to the random orientation of self-propulsion. The diffusivity is more in the case of active particles than in passive ones.  
In the presence of particle interaction, as $v_0$ increases, the force due to the self-propelled nature of the particles dominates over the oscillatory force. In this limit, particles of lighter mass move with higher velocity, exhibiting a more pronounced peak.    

\section{Discussion}
\label{sec:discussion}

From the above observation, it is evident that $\langle v \rangle$ shows a bimodal behaviour in the presence of interaction between the particles. 
Based on the observations, we propose empirical relations to extract the optimal mass $m_{op}^h$ ($m$ corresponds to the peak value of $\langle v \rangle$) at moderate mass values as a function of the parameters of the oscillatory force $a, \, \omega$ and the interaction strength $k$ as $m_{op}^h \sim (a\,\omega^2\,k^2)^{-0.4}$. Note that $m_{op}^h$ is not much influenced by other parameters, i.e., $v_0$ and $\epsilon$.   
Similarly, the optimal mass in the lower mass region is $m_{op}^{l}\sim (v_0^4\,/k^3)^{0.8}$, with a condition $k \ge 350$. 
Note that in the absence of $k$, there is no additional peak in the lower $m$ region.
By changing the asymmetry of the channel geometry, defined by the parameter $\epsilon$, the peak height in $\langle v \rangle$ with respect to $m$ changes but not the peak position significantly.
These observations are very useful in sorting out the particles of various masses. 
The observed mass-based separation mechanism can be studied experimentally for active particles of various
masses diffusing in an asymmetric channel, which can be
prepared by microprinting on a substrate \cite{Mahmud2009}. 
To have an estimate in real units, which is very useful for
the experimentalists, the characteristic values of the friction
coefficient and characteristic diffusion time for the particles
in water moving in a triangular channel with aspect ratio $\epsilon = 0.1$ 
and period length $L \sim 10 \, \mu m$ at room temperature $(T \sim 300 K)$ 
are given by $\eta \sim 2 \times 10^{-3} mg/s$ \cite{Cussler2008} and $\tau \sim 50 \,s$, respectively \cite{Haenggi2009, Matthias2003}. Particles having mass around $M_{op}^{h} \sim 38 \times 10^{-3} mg$ and subject to parameters $A \sim 0.25 \,fN$, $v_0$ : $v_0 \sim 0.207 \,\mu m/s$, $k \sim 144 \,fN$  and $\Omega \sim 0.0002 \, s^{-1}$ move at an average velocity of about $0.13 \,\mu m/s$, which is higher than that of other particles. Similarly, Particles having moderate mass around $M_{op}^{l} \sim 0.193 \times 10^{-3} mg$ move at an average velocity of about $0.12 \, \mu m/s$.
It is expected that these results will motivate the experimentalists to design lab-on-a-chip devices
for separating active particles, nano-and micro-particles, proteins, organelles,
and cells based on their mass.

\section{CONCLUSION}
\label{sec:conclusions}

In this work, we have studied the diffusive behaviour of interacting active particles with mass $m$ in an asymmetric channel. 
The confinement of the channel significantly controls the diffusive behaviour of the particles. As the considered channel is asymmetric in shape, particles exhibit rectification. 
Without particle interaction, the mean velocity $\langle v \rangle$ and the corresponding effective diffusion $D_{eff}$ show a single peak at moderate $m$. 
However, due to the interaction between the particles, $\langle v \rangle$ exhibits a bimodal behaviour. It indicates that particles of moderate and lighter particles can exhibit higher rectification than the rest. 
By altering the strength of interaction, self-propelled velocity, and the parameters of the oscillatory force, we can selectively extract the particles of either lower or moderate mass. 
We have proposed empirical relations, $m_{op}^{h} \sim (a \,\omega^2\,k^2)^{-0.4}$ for moderate mass and $m_{op}^{l}\sim (v_0^4\,/k^3)^{0.8}$ for lower mass with the condition that the particle interaction should be strong enough ($k \ge 350$). 
Note that the asymmetry of the channel does not play a vital role in $m_{op}^{h}$ or $m_{op}^{l}$. 
These findings are helpful in separating the nano- and micro-particles, proteins, organelles, and cells.

\section{ACKNOWLEDGMENT}
This work was supported by the Indian Institute of Technology Kharagpur.


%

\end{document}